\documentclass[a4paper,oneside]{saip}
\usepackage{helvet}          
\usepackage{times}           
\usepackage{courierten}   
\usepackage[T1]{fontenc} 

\usepackage{graphicx}  
\usepackage{url}
\usepackage{amsmath}
\usepackage{geometry} 
\begin{document}

\title{Implementing $F(T)$ Gravity in Boltzmann Codes: A Framework for Power Spectrum Computation}

\author{Robert Rugg$^{1}$, Shambel Akalu$^{1,2}$ and Amare Abebe$^{1,3}$}

\affil{$^1$Centre for Space Research, North-West University, Potchefstroom 2531, South Africa}
\affil{$^2$Department of Physics,  Wolkite University, Wolkite, Ethiopia}
\affil{$^3$National Institute for Theoretical and Computational Sciences (NITheCS), Potchefstroom 2520, South Africa}

\email{\url{31770312@mynwu.ac.za}}

\begin{abstract}
This work investigates the non-linearity of the power-law model of $F(T)$ gravity, highlighting the Boltzmann solver CLASS’s inability to handle nonlinear models. A second-order Taylor expansion is applied to the nonlinear field equations as a workaround, under the assumption that the extra degree of freedom ($n$), which quantifies deviations from the currently favored cosmological model ($\Lambda$CDM), remains small enough to preserve the key properties of the $\Lambda$CDM model. The Taylor expansion is supported by supernova data that indicate $n \leq 0.05$, allowing the power spectrum to be accurately computed within CLASS with a negligible truncation error for these values of $n$.
\end{abstract}

\section{Introduction}

High-redshift observations, notably those of the High-Z supernova search team, have revealed that the universe is undergoing accelerated expansion \cite{riessObservationalEvidenceSupernovae1998}. This behaviour is attributed to a negative pressure component commonly referred to as dark energy. The currently favoured cosmological model, $\Lambda$CDM, attributes this acceleration to a cosmological constant \cite{peeblesCosmologicalConstantDark2003}. However, it faces challenges such as the Hubble tension \cite{simpsonLambdaCDMHowHubble2025} and the fine-tuning problem \cite{huCosmicMicrowaveBackground2002}.
\\ \\
Alternative explanations for dark energy include models involving scalar fields, such as phantom fields with negative kinetic energy, canonical quintessence models, and interacting dark energy fluids \cite{valentinoCosmoVerseWhitePaper2025}. A more direct approach involves modifying gravity itself by generalizing the gravitational action to include arbitrary functions of geometric scalars, such as $f(Q)$ \cite{sahluConstrainingViscousfluidModels2025}, $f(R)$, and $f(T)$ \cite{caiTeleparallelGravityCosmology2016a}.
\\ \\
In this work, we focus on the torsion-based formulation of gravity, where the Weitzenböck connection replaces curvature with torsion. This framework utilises the tetrad formalism \cite{caiTeleparallelGravityCosmology2016a} and constructs the Lagrangian density from the torsion scalar $T$, preserving local Lorentz and coordinate invariance \cite{hayashiNewGeneralRelativity1979}. Of particular interest is the generalised form $f(T)$, where the action becomes a function of $T$ rather than $T$ itself \cite{linderEinsteinsOtherGravity2010}. This generalisation, referred to as $f(T)$ gravity, extends the Teleparallel Equivalent of General Relativity (TEGR), and introduces second-order field equations with richer gravitational dynamics.
\\ \\
This study examines the power-law model of $f(T)$ gravity proposed in \cite{linderEinsteinsOtherGravity2010}, which has proven advantageous for data fitting. We adapt the CLASS Boltzmann solver \cite{diegoblasCosmicLinearAnisotropy2011} to incorporate this model and analyse its implications for the cosmic microwave background (CMB) power spectrum. The structure of this paper is as follows. Section 2 introduces the background cosmology of $f(T)$ gravity and outlines the numerical methods used to adapt CLASS. Section 3 covers scalar perturbations and presents the resulting power spectrum. Section 4 concludes the study.

\section{Background cosmology of f(T) gravity}
For this work, Greek indices $\mu,\nu,...$ are associated with spacetime coordinates (0,1,2,3) while capital Latin indices A,B,.. are associated with tangent spacetime coordinates. Lower case Greek and Latin indices are then associated with spatial and tangent spatial coordinates (1,2,3), respectively. In Teleparallel gravity, the underlying geometry used to describe spacetime is known as the tetrad-spin connection and describes the Lorentzian metric $g_{\mu \nu} = \eta_{AB} e^{A}_{\mu}e^{B}_{\nu}$, defining the Minkowski metric as $\eta_{AB} = \text{diag}(1,-1,-1,-1)$
\cite{bahamondeTeleparallelGravityTheory2023}. To assume the non-degenerate characteristic, the inverse tetrad ($E_{A}^{\mu}$) must satisfy $E_{A}^{\mu}e^{A}_{\mu} = \delta^{\mu}_{\nu}$. The affine connection as in the torsion scenario is then described as the Weitzenböck connection:
\begin{equation}
    \hat{\Gamma}^{\rho}_{\mu \nu} = E_{A}^{\rho}(\partial_\nu e^{A}_{\mu} + \hat{w}^{A}_{B\nu}e^{B}_{\nu}),
\end{equation}
where $\hat{w}^{A}_{B\nu}$ is defined as the spin connection, which in most background \cite{bahamondeTeleparallelGravityTheory2023} and perturbative \cite{golovnevPerturbationsFTCosmology2020} cases, the inclusion of the spin connection amounts to the same equations of the pure tetrad formalism \cite{krssakCovariantFormulationFT2016} and only restores Lorentz invariance. In general relativity, curvature was associated with the Levi-Civita connection; however, in Teleparallel gravity, the curvatureless property would result in a zero Riemann tensor. It is possible to relate the two theories through a general affine connection where the difference between the Weitzenböck and Levi-Civita connections is highlighted in the contorsion tensor ($\hat{K}^{\rho}_{\mu \nu}$) such that:
\begin{equation}
    \begin{split}
        &\hat{K}^{\rho}_{\mu \nu} = \frac{1}{2}(\hat{T}_{\nu }^{\mu}{}_{\rho} + \hat{T}_{\rho} ^{\mu}{}_{\nu} - \hat{T}^{\mu}_{\rho \nu}) \\
        &S_{\rho}^{\mu \nu} = \frac{1}{2}(K^{\mu \nu}_{\rho} + \delta^{\mu}_{\rho}T^{\alpha \nu}_{\alpha}  - \delta^{\nu}_{\rho}T^{\alpha \mu}_{\alpha}),\\
    \end{split}
\end{equation}
where the torsion scalar $T = S_{\rho}^{\mu \nu}T^{\rho}_{\mu \nu}$ is the product of the torsion tensor and the super potential $S_{\rho}^{\mu \nu}$. This is similar to the General Relativity (GR) case where the Riemann tensor can be related to the Ricci scalar.
\\ \\
In the TEGR case, the action is calculated by the Lagrangian density $T$; however, in the case of generalised Teleparallel gravity, the torsion scalar in the Lagrangian is replaced by an arbitrary function $f(T) = -T + F(T)$. It is clear that the case of $\Lambda$CDM is recoverable in $F_T(T) = F_{TT}(T) = 0$ (first- and second-order derivatives of the function $F(T)$ with respect to the torsion scalar) implying a constant value $F(T)$ as in the constant case of $\Lambda$CDM. The generalised function then reduces to the TEGR case, which is equivalent to the GR case \cite{weinbergCosmology2008}. The action for the generalised case then reads:
\begin{equation}
    \mathcal{S} = \int d^4x |e| \left[ \frac{f(T)}{16\pi G} + L^{m}\right],
\end{equation}
such that $|e| = det(e^A_\mu) = \sqrt{-g}$ and $L^m$ is the matter Lagrangian. Assuming a spatially flat Friedmann-Robertson-Walker (FRW) metric such that the vierbein takes the form $e^{A}_{\mu} = \text{diag}(1,a,a,a)$ with a perfect matter fluid, the modified Friedmann equations can be written in the form:
\begin{equation}
    \begin{split}
    &H^2  = -\kappa^2 \rho  + \frac{1}{6}F(T) - \frac{1}{3}TF_T(T),\\ 
     &\dot{H} = - \frac{3}{2}H^2 - \frac{1}{2}\kappa^2 P + \frac{1}{2}(6H^2F_T + 2\dot{H}F_T + 24H^2\dot{H}F_{TT} +\frac{1}{2} F(T)),\\
    \end{split}
    \label{eq:Friedmann}
\end{equation}
where $\rho$ and $P$ refer to the energy density and pressure associated with matter and radiation, with $T = 6H^2$. An effective dark energy density and pressure:
\begin{equation}
    \begin{split}
    &\rho_{\rm eff} = \frac{1}{\kappa^2}(\frac{1}{2}F(T) - TF_T(T)),\\ 
     &P_{\rm eff} = \frac{1}{\kappa}(6H^2F_T + 2\dot{H}F_T + 24H^2\dot{H}F_{TT} +\frac{1}{2} F(T)),\\
    \end{split}
\end{equation}
can be deduced from the above equations, where an effective equation of state naturally arises as $w_{\rm eff} = \frac{P_{\rm eff}}{\rho_{\rm eff}}$. This differs from the $\Lambda$CDM case where $w_{\rm eff} = -1$, assuming a constant "dark energy", while this modified version of the TEGR case clearly indicates a dynamical dark energy. 
\section{The power-law model}
\begin{figure}[ht]
\centering
\includegraphics[width=0.5\textwidth]{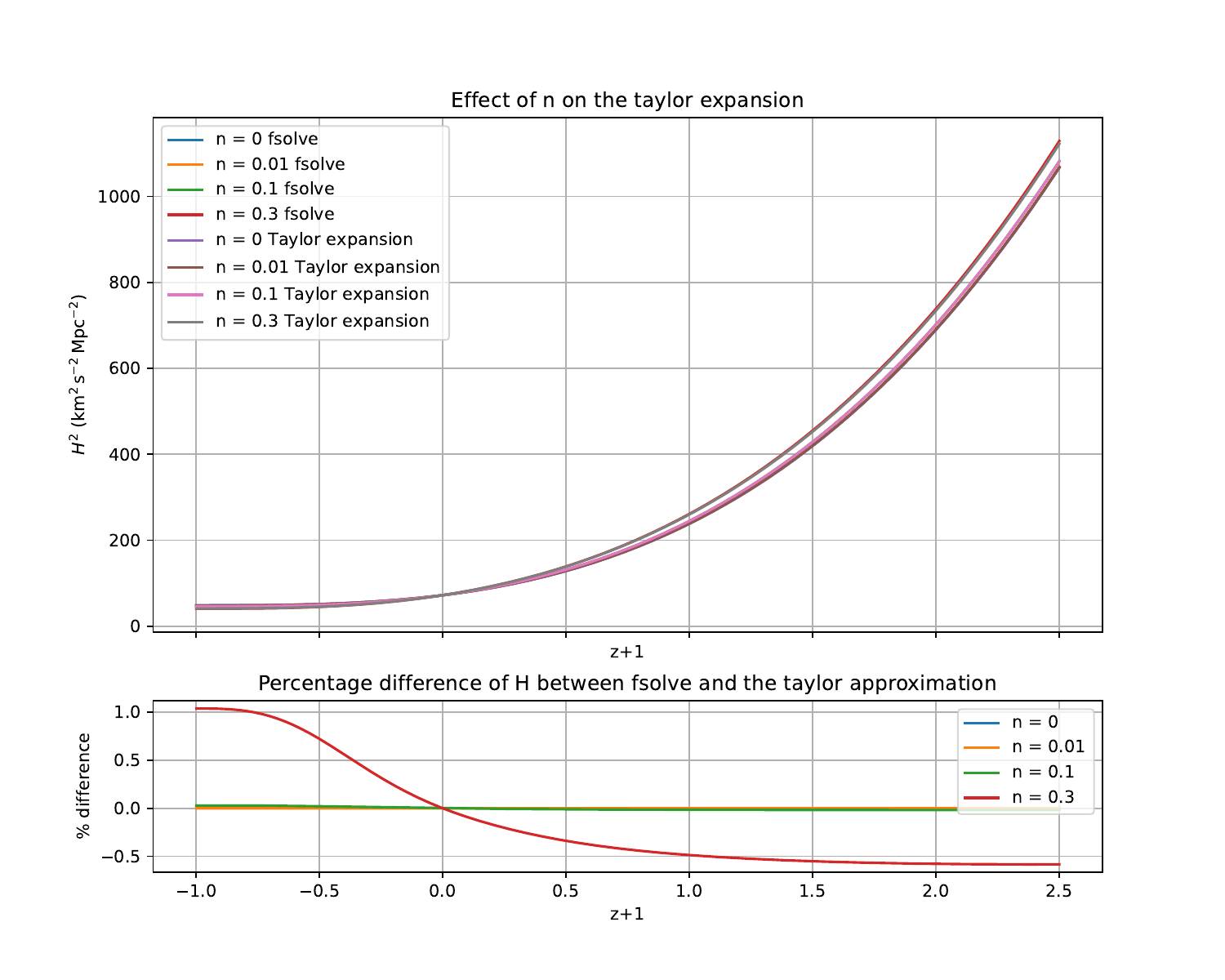}
\caption{\label{fig-Percentage} (Top) The Hubble parameter plotted for various values of $n$ using both the second-order Taylor expansion method and \texttt{scipy}'s \texttt{fsolve} as a function of $z$. (Bottom) Percentage difference between the second order Taylor expansion method and \texttt{scipy}'s \texttt{fsolve} at specific values of $n$ as a function of $z$ for $H^2$.} 
\end{figure}

This paper follows works in \cite{nesserisViableFTModels2013}, where the $f(T)$ -ansatz (power law) $F(T) = \alpha(-T)^n$ \cite{bengocheaDarkTorsionCosmic2009} is assumed. If one assumes the perfect fluid, then it is possible to solve for $\rho_{\rm eff}$, such that $\dot{\rho}_{\rm eff} + 3H(1 + w_{\rm eff}) = 0$, where $w_{\rm eff} = P_{\rm eff}/\rho_{\rm eff}$, resulting in a more convenient dynamical dark energy written in dimensionless parameters equivalent to $\alpha(-T)^n$:
\begin{equation}
    \rho_{\rm eff} = \rho_{eff,0}E^{n}(z),
    \label{eq:DarkEnergy}
\end{equation}
where the subscript $_{eff,0}$ refers to the density at redshift $z = 0$ and $E(z) = \frac{H(z)}{H_0}$. It is clear from this term that the dynamical dark energy returns to its non-dynamical form as a constant in the case $n = 0$. This nonlinear form arising from the power law model can be solved numerically. It is of great interest to find the power spectrum of the model, that will be generated using the CLASS Boltzman equation \cite{diegoblasCosmicLinearAnisotropy2011}. CLASS, is, however, ill equipped to handle non-linear forms such as the power law, and hence a less rigorous numerical approach of a second-order Taylor expansion as in \cite{nesserisViableFTModels2013}, is taken around $n = 0$ with \ref{eq:DarkEnergy}, such that:
\begin{equation}
    \begin{split}
    &E^2(z,n) = E_\Lambda^2(z,0) + \frac{dE^2(z,n)}{dn}\Bigg|_{n=0} + \frac{d^2E^2(z,n)}{dn^2}\Bigg|_{n=0} + \text{ ...}, \\
    &E^2(z,n) = E_\Lambda^2(z) + \Omega_{\rm eff}\ln{E^2_{\Lambda}} n + \left(\Omega_{\rm eff}\ln^2{E^2_\Lambda} +
         \frac{2\Omega_{\rm eff}\ln{E^2_{\Lambda}}}{E^2_{\Lambda}}\right)\frac{n^2}{2} \text{ ...},\\ 
    \end{split}
    \label{eq:Taylor}
\end{equation}
where $E_\Lambda = \sqrt{\Omega_{m,0}(z+1)^3 + \Omega_{r,0}(z+1)^4 + \Omega_\Lambda}$ refers to the $\Lambda$CDM Friedmann equation with the subscripts m,r and $\Lambda$ being matter, radiation, and dark energy, respectively. $\Omega$ is introduced as a dimensionless density corresponding to their respective components, denoted by i such that $\Omega_{i} = \frac{8\pi G}{3H^2(t)} \rho_{i}$. In addition, $n \ll 1$, as there is no expectation of differing significantly from $\Lambda$CDM, which would result in inadequate data fitting as seen in the figure below. In addition, the assumption $n\ll 1$ is supported by supernova results for f(T) \cite{kumarNewCosmologicalConstraints2023}, who reported a small constrained value of 0.044. For this small value, the truncation error from the second-order Taylor expansion is estimated to be negligible.
\\ \\
The second approximation technique was computed with \texttt{scipy}'s \texttt{fsolve} \cite{virtanenSciPy10Fundamental2020} in Python where no assumptions were made about the value of $n$. The curves labeled "fsolve" in Fig. \ref{fig-Percentage} correspond to this full numerical solution, while the curves labeled "Taylor expansion" show the second-order Taylor expansion. It is clear from Fig. \ref{fig-Percentage}, that as the value of $n$ increases, the percentage difference between the two methods increases significantly. This will clearly bias the results towards a small value of $n$, when the Taylor expansion is incorporated into the CLASS Boltzman solver for CMB data.

\subsection{Newtonian-gauge perturbations in F(T) gravity}
\begin{figure}[ht]
\centering
\includegraphics[width=0.6\textwidth]{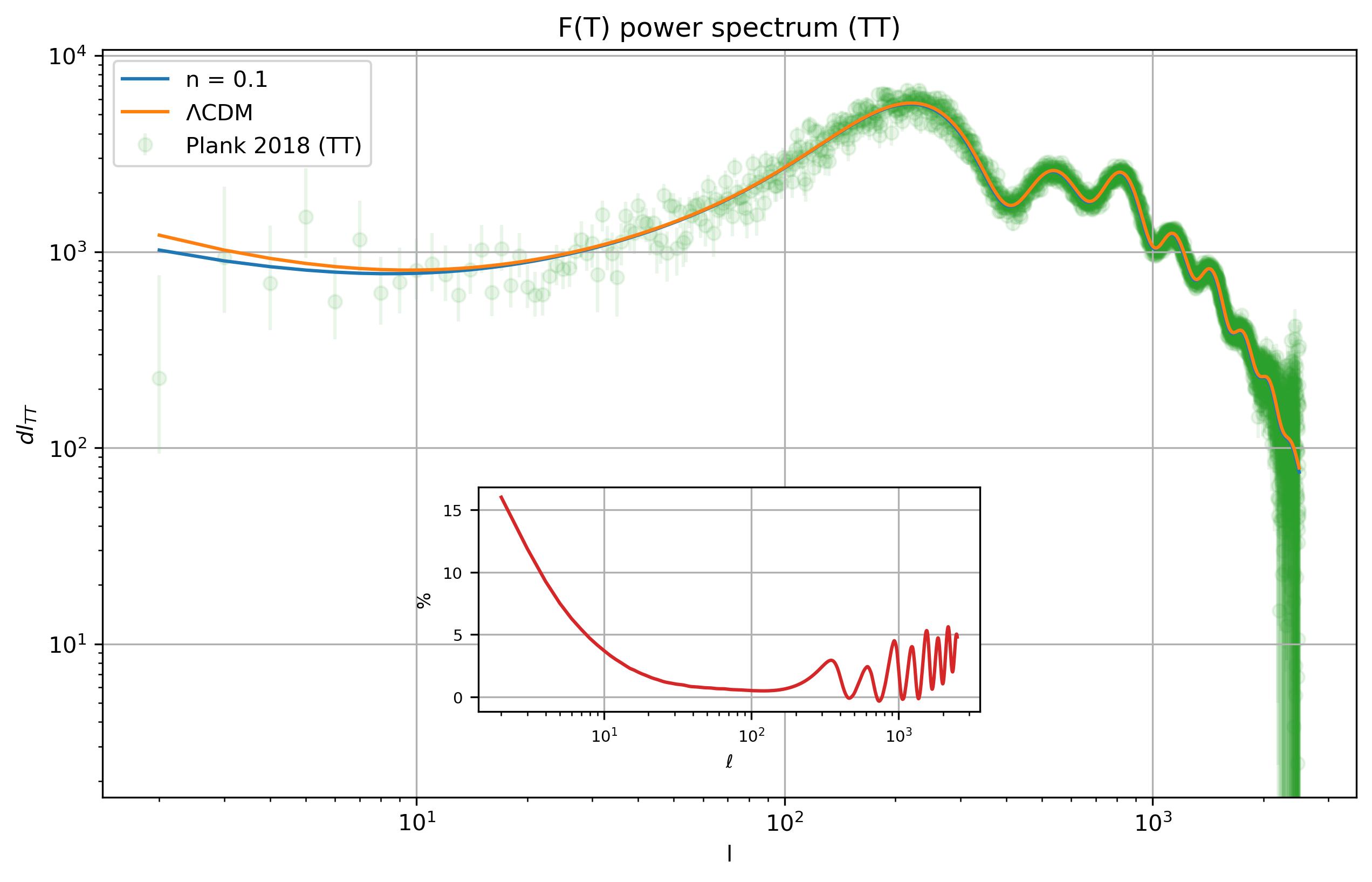}
\caption{\label{fig-Power} (Top) The power spectrum for the $\Lambda$CDM and $F(T)$ power law case plotted against temperature mapping data form the Planck mission for different multipole moments $l$. (Bottom) The percentage difference between the $\Lambda$CDM and the $F(T)$ power-law model over different multipole moments $l$.} 
\end{figure}

Following the adjustments made in the CLASS Boltzman solver for the background cosmology, perturbations play an essential role in theoretical anisotropies. The following derivation is followed from \cite{chenCosmologicalPerturbationsFT2011} in which the metric formalism is reconstructed using a Newtonian gauge such that scalar modes $\phi$ and $\psi$ are introduced as functions of spacetime coordinates where:
\begin{equation}
    ds^2 = (1 + 2\psi)dt^2 - a^2(1-2\phi)\delta_{ij}dx^{i}dx^{j}.
\end{equation}
Since the Boltzmann solver CLASS is set up in the more convenient Fourier space and conformal time ($\mathcal{H} = H/(1+z)$), the perturbed equations were converted accordingly using the mode expansion where $k$ is introduced as the co-moving wave number related to the co-moving wavelength ($k = \frac{2\pi}{\lambda}$), such that:
\begin{equation}
    \begin{split}
        &\phi(t,\boldsymbol{X}) = \frac{1}{(2\pi)^{\frac{3}{2}}} \int dk^3 \tilde{\phi}_k(t)e^{i\boldsymbol{k}\cdot\boldsymbol{x}},\\
        &\psi(t,\boldsymbol{X}) = \frac{1}{(2\pi)^{\frac{3}{2}}} \int dk^3 \tilde{\psi}_k(t)e^{i\boldsymbol{k}\cdot\boldsymbol{x}}, \\
    \end{split}
\end{equation}
where $\phi$ and $\psi$ are the scalar metric perturbations in the Newtonian gauge such that $T_1 = 12H(H\psi + \dot{\phi})$. The momentum constraint equation and the shear equation are sufficient to overlay the effects of perturbations on observables such as the CMB \cite{dioCLASSgalCodeRelativistic2013} with $\sigma$ and ${u}_k$ the shear and velocity perturbation of the matter fluid, respectively, such that:\label{eq:einstein}
\begin{equation*}
    \begin{split}
        \dot{\tilde{\phi}}_k  &= -\mathcal{H} \tilde{\psi}_k + \frac{4\pi (\rho_m + p_m) \tilde{u}_k}{(12(\frac{\mathcal{H}}{a})^2   F''_0 - (1 + F'_0))}  \\
        \tilde{\psi}_k &=  \tilde{\phi}_k - \frac{4\pi G a^2k^2}{1+F'_0} \sigma.\\
    \end{split}
\end{equation*}
The equations listed in \ref{eq:Friedmann}, contain all relevant background evolutions for power-spectrum analysis. The Taylor expansion in Equation \ref{eq:Taylor} are used to overcome CLASS's inability to handle non-linear Friedmann equations. The negative pressure associated with the cosmological constant in the $\Lambda$CDM model was turned off and simulated with the modified Friedmann Equation \ref{eq:Friedmann}. This paper will mainly focus on temperature mapping with CMB$_{TT}$ data and Figure \ref{fig-Power} below is a representation of the modified Friedmann equations using the $F(T)$ power-law model.
\\
It is clear from Figure \ref{fig-Power} a large initial difference comes from the dynamical dark energy in the power-law case that has direct implications on gravitational potentials before recombination. Thus, the dynamical dark energy has a direct impact on the frozen photon-baryon fluid. In addition, the shape and peaks of each acoustic waves are directly affected, and most of the implications are associated with the slightly higher percentage difference prevalent in the tail, indicating that $F(T)$ has direct consequences on the baryon density or matter density at recombination. The percentage difference also points towards stronger peak distortion than shape distortion, again indicating that $F(T)$ has an impact on the baryon-photon interactions.
\begin{figure}[ht]
\centering
\includegraphics[width=0.6\textwidth]{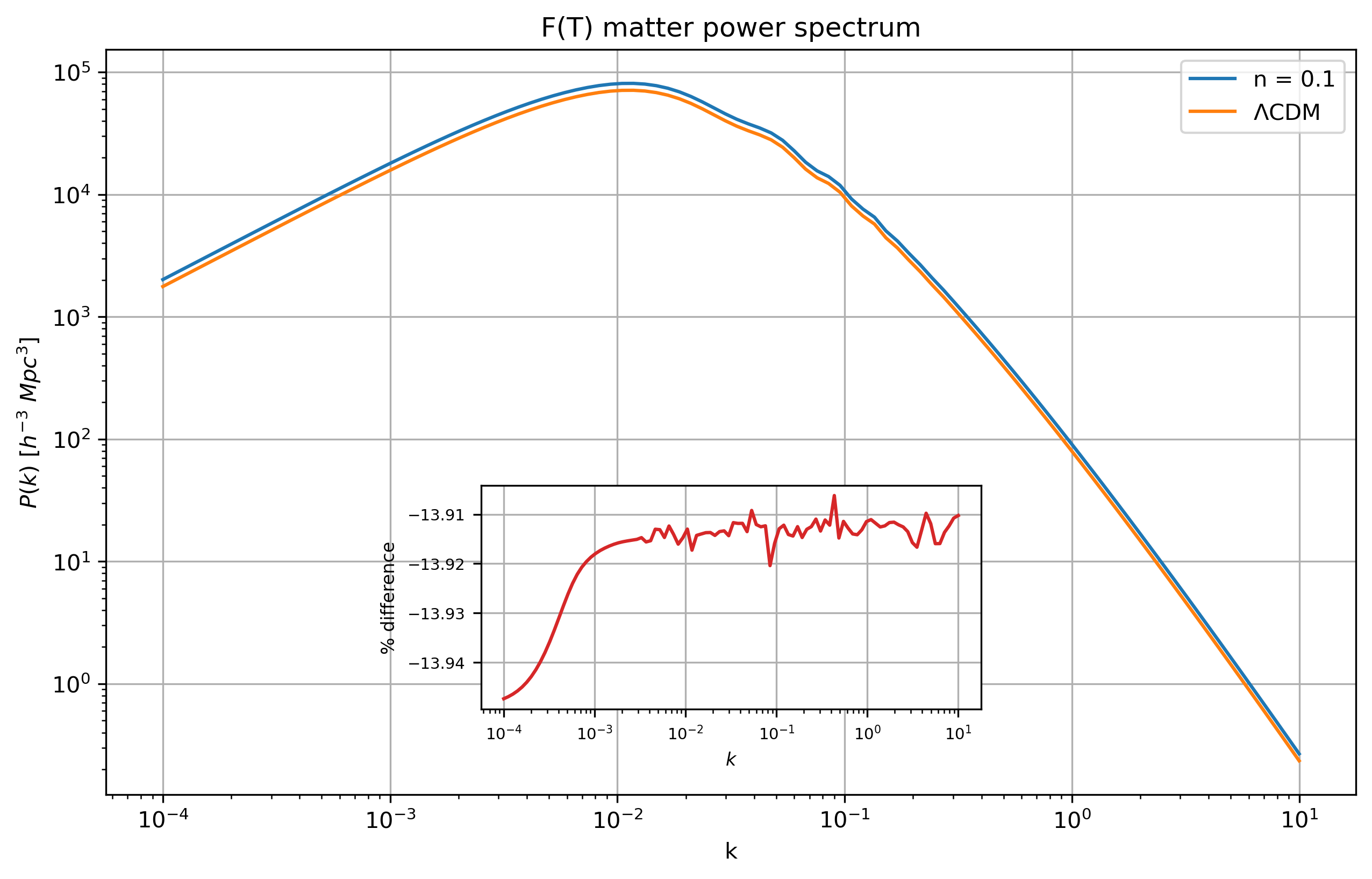}
\caption{\label{fig-Matter} The matter power spectrum for the $\Lambda$CDM and the $F(T)$ power-law case plotted for different wavenumbers ($k$).  The percentage difference between the $\Lambda$CDM and the $F(T)$ power-law model with $n = 0.1$ is indicated in the subplot within.} 
\end{figure}
\\ \\
Figure \ref{fig-Matter} shows that $F(T)$ follows a very similar trend to $\Lambda$CDM. However, at $n = 0.1$, $F(T)$ gravity explores higher amplitudes, indicating a larger clustering of matter. It should also be noted that $n$ is chosen as $n=0.1$, but realistically expected to be lower. This would indicate a much lower difference, which is crucial as galaxy surveys are extremely sensitive regarding the matter power spectrum.

\section{Conclusions}
The second-order Taylor expansion serves as a viable approach to solve the non-linearity issues faced in CLASS. The difference between \texttt{scipy}'s \texttt{fsolve} shows a difference compared to the Taylor expansion approach of $1\%$ at $n = 0.1$. It is well-reported that a small value of $n$ is preferred as data surveys are extremely sensitive to the power spectrum. CLASS effectively generates the matter power spectrum and temperature fluctuations, which will perfectly allow for constraining of the power-law model. There is a sense of bias to this result, as $n$ is assumed to be small in nature, but any large value of $n$ would result in large likelihoods due to its large difference from the $\Lambda$CDM model which fits the data extremely well.
\\ \\
Using the second-order Taylor expansion effectively highlights the effect that a dynamical effective equation of state has on the power spectrum, producing a large difference at low multipoles. The effects of $F(T)$ gravity are largely observed in the shape and peaks of each acoustic wave, where the shape of the acoustic wave was slightly altered. Moreover, there appear to be larger effects on the damping tail, indicating the effect on matter densities at recombination. Finally, the matter power spectrum indicates a trend similar to $\Lambda$CDM at lower values of $n$, but indicates slightly more matter clustering. 

\pagebreak

\bibliographystyle{IEEEtran}
\bibliography{SAIP_proceedings/Masters}

\end{document}